\documentclass[aps,pra,reprint,superscriptaddress,amsmath,showpacs,amssymb]{revtex4-1}
\usepackage{bm}
\usepackage[dvips,final]{graphicx}
\usepackage{amsmath,amssymb}

\newcommand{\be}{\begin{equation}}
\newcommand{\ee}{\end{equation}}
\newcommand{\BM}{\begin{pmatrix}}
\newcommand{\EM}{\end{pmatrix}}

\renewcommand{\d}{\dagger}
\renewcommand{\phi}{\varphi}

\newcommand{\bra}[1]{\bigl\langle #1 \bigr|}
\newcommand{\ket}[1]{\bigl| #1 \bigr\rangle}
\newcommand{\braket}[2]{\bigl\langle #1 \big| #2 \bigr\rangle}
\newcommand{\brasub}{{\!\!\phantom{\big\langle}}}

\newcommand{\intx}{\int\!d^3x\;}
\newcommand{\pfrac}[2]{\frac{\partial #1}{\partial #2}}
\newcommand{\phiex}{\phi_{\mathrm{ex}}}
\newcommand{\avg}[1]{\bra0 #1 \ket0}
\newcommand{\SEC}[1]{\section{#1}}
\bmdefine{\bx}{x}
\bmdefine{\by}{y}
\bmdefine{\bz}{z}

\begin{document}
\title{Formulation for zero mode of Bose-Einstein condensate beyond Bogoliubov approximation}

\author{Y.~Nakamura}
\email{nakamura@aoni.waseda.jp}
\affiliation{Department of Electronic and Photonic Systems, Waseda
University, Tokyo 169-8555, Japan} 
\author{J.~Takahashi}
\email{phyco-sevenface@asagi.waseda.jp}
\affiliation{Department of Electronic and Photonic Systems, Waseda
University, Tokyo 169-8555, Japan} 
\author{Y.~Yamanaka}
\email{yamanaka@waseda.jp}
\affiliation{Department of Electronic and Photonic Systems, Waseda
University, Tokyo 169-8555, Japan} 

\date{\today}

\begin{abstract}
It is shown for the Bose-Einstein condensate of cold atomic system that the new unperturbed Hamiltonian, 
which includes not only the first and second powers of the zero mode operators 
but also the higher ones,
determines a unique and stationary vacuum at zero temperature.
From the standpoint of quantum field theory,
it is done in a consistent manner that the canonical commutation relation of the field operator is
kept. In this formulation, the condensate phase does not diffuse and 
is robust against the quantum fluctuation of the zero mode. The standard deviation for the phase
operator depends on the 
condensed atom number with the exponent of $-1/3$, which is universal for both
homogeneous and inhomogeneous systems. 
\end{abstract}



\maketitle

\section{Introduction}
Since the realization of Bose-Einstein condensate (BEC) 
in cold atomic systems \cite{Cornell,Ketterle,Bradley}, 
it has been offering challenging subjects in theoretical foundations of quantum many-body problem. 
The theoretical description of BEC uses the complex order parameter or macroscopic wave function
which is subject to the Gross-Pitaevskii (GP) equation, and is
in very good agreement with the experiments at lower temperatures \cite{Dalfovo}, as long as
atomic interaction is weak and both quantum and thermal fluctuations can be neglected.
The experimental observation of the interference between two condensates \cite{interference} 
indicates that the order parameter of each condensate has a definite phase.
The theoretical calculation based on the order parameter reproduces 
the interference fringe accurately \cite{Rohrl}. 

Quantum field theory, as the most fundamental one of quantum many-body problem, provides us
with a clear and sound interpretation of the condensation: It is the ordered state associated with 
the spontaneous breakdown of the global gauge symmetry.  At the same time the zero mode 
must exist according to the Nambu-Goldstone theorem \cite{NGtheorem1,NGtheorem2}, implying that
it plays a crucial role in creating and retaining the ordered state. 
However, it is often, typically in the Bogoliubov approximation, neglected
 in formulating the quantum fluctuation at the sacrifice of the theoretical consistency.
This is mainly because of its infrared singular property which is intractable and 
of a naive and groundless expectation that it does not affect the system very much.
In this paper, we take full account of the zero mode from the standpoint of quantum field theory,
in which the order parameter at zero temperature is given by the vacuum expectation of 
the field operator. The calculational scheme of quantum field theory
is constructed in the interaction picture 
with an unperturbed Hamiltonian which contains the first and second powers 
of the field operator only but none of higher ones. 
Both of the zero and excitation modes on 
the condensate are described by the Bogoliubov-de Gennes (BdG) equation
 \cite{Bogoliubov,deGennes,Fetter}, and the unperturbed field operator is expanded in the BdG complete set.
While the unperturbed Hamiltonian of the excitation mode sector is diagonalized, that of the zero mode one 
is not \cite{Lewenstein,Matsumoto2,Mine}. 
Rather the dynamics of the zero mode should be represented by a pair of 
quantum coordinate $Q$ and momentum $P$, and the corresponding Hamiltonian is one of a free particle.
Then the vacuum including the zero mode sector can not be determined for certain, because
 a stationary ground state does not exist. To overcome this problem, 
Lewenstein and You have introduced a new expansion of the field operator in which $Q$ becomes 
a phase operator of the condensate and concluded a phase diffusion,  growing quadratically in time \cite{Lewenstein}. We however
note that their new field operator breaks the canonical commutation relation which is the very foundation 
of quantum field theory. Moreover, the predicted phase diffusion has not been observed
experimentally, and several experiments are consistent with no phase diffusion \cite{Hall,Greiner,Gati}. 
The treatment of the zero mode is still an open question. 

In what follows, we consider a Bose--Einstein condensed system of weakly interacting atoms 
at zero temperature.  Our proposition is that all the terms of the total Hamiltonian, 
consisting only of $Q$ and $P$, are taken in the unperturbed Hamiltonian. Then we naturally 
obtain a unique vacuum 
which is a stationary ground state of the zero mode sector and causes no infrared divergence, 
while the canonical commutation relation is not violated. We conclude from the evaluation of 
the variance in phase that no phase diffusion occurs
and that the standard deviation of the phase decreases with the exponent 
$-1/3$ as the condensate number increases. 
It is shown that the exponent is universal and independent of whether the system is homogeneous or not.

\section{Ordinary Formulation and Dilemma of Zero Mode}
Let us briefly sketch the ordinary formulation for the system 
at zero temperature, and see the dilemma of the zero mode mentioned above.
 We start with the total Hamiltonian,
\be
	H = \intx  \left[ \psi^\d \left(-\frac{\nabla^2}{2m} + V_{\mathrm{ex}} - \mu\right) \psi
  + \frac{g}{2} \psi^\d\psi^\d\psi\psi \right] \,,
\ee
where $m$, $V_{\mathrm{ex}}$, $\mu$, and $g$ represent the mass of an atom, the confinement potential, the chemical potential, 
and the coupling constant, respectively, and $\hbar$ is set to be unity. The bosonic field 
operator $\psi$ obeys the canonical commutation relations ($x=({\bm x},t)$)
\be
	\bigl[ \psi(x) , \psi^\d(x') \bigr]_{t=t'} = \delta(\bx-\bx') \,,\quad
	\bigl[ \psi(x) , \psi(x') \bigr]_{t=t'} = 0 \,,
\ee
and is divided into a classical part $\xi$ and an operator $\phi$ on the 
criterion $\bra0 \phi \ket0 = 0$. Note that the vacuum $\ket0$ is not specified yet and should 
be determined self-consistently. The total Hamiltonian is rewritten in terms of $\phi$ as
\be
	H = H_1 + H_2 + H_3 + H_4 \,,
\ee
where
\begin{align}
	H_1 &= \intx \left[ (\phi+\phi^\d)(h_0 - \mu + g\xi^2)\xi \right] \,,\\
	H_2 &= \intx \left[ \phi^\d \mathcal{L} \phi + \frac12\phi\mathcal{M}\phi + \frac12\phi^\d\mathcal{M}\phi^\d \right] \,,\label{eq:defH2}\\
	H_3 &= g\intx \xi (\phi^\d\phi\phi + \phi^\d\phi^\d\phi ) \,,\\
	H_4 &= \frac{g}{2}\intx \phi^\d\phi^\d\phi\phi\,,
\end{align}
with 
$h_0 = -\nabla^2/2m+V_{\mathrm{ex}}\,,$ 
$\mathcal{L} = h_0 -\mu + 2g\xi^2 \,,$ 
$\mathcal{M} = g\xi^2\,$.
Here the order parameter $\xi$ is taken to be real for simplicity.

On the premise of small $\phi$, the customary step is 
to choose $H_1 + H_2$ as the unperturbed Hamiltonian except for the renormalization counter terms
 in the interaction picture. From $\bra0 \phi \ket0=0$ for a time-independent vacuum $\ket{0}$ and
any $t$ follows
\be	\label{eq:timederivation}
	i\partial_t \bra0 \phi \ket0 = \bra0 [\phi, H_1 + H_2] \ket0 = 0 \,.
\ee
It implies $H_1=0$ and therefore the GP equation at the leading order \cite{GP},
\be \label{eq:GP}
	(h_0 - \mu + g\xi^2)\xi=0\,.
\ee 
In an attempt to diagonalize $H_2$ \cite{Lewenstein,Matsumoto2,Mine}, we introduce the BdG equation  
$
	T \by_\ell = \omega_\ell \by_\ell\,
$
with the doublet notations,  
\be
	T = \BM \mathcal{L} & \mathcal{M} \\ -\mathcal{M} & -\mathcal{L} \EM  \,,\qquad 
	\by_\ell = \BM u_\ell \\ v_\ell  \EM\,.
\ee
Due to the non-hermiticity of $T$, $\omega_\ell$ can be complex in general. The diagonalization of 
 the complex mode part is also a subject to be settled \cite{Mine2007}. We restrict ourselves only
to the real eigenvalues below. Because the global phase symmetry is spontaneously broken, there is a 
eigenfunction belonging to a zero eigenvalue, {\it i.e.} $T \by_0=0\,$ with $\by_0 = (\xi ,\; -\xi)^t\,$, and an additional 
adjoint function $\by_{-1} = ( \eta ,\; \eta)^t$ has to be introduced for the completeness, where 
$\eta = \pfrac{\xi}{N_0}$ and $N_0 = \intx \xi^2\,$. We adopt the following linear expansion of 
$\phi(x)$ by the BdG complete set, 
\begin{align} \label{eq:phi_expansion}
	\phi(x) &= -iQ(t)\xi(\bx) + P(t) \eta(\bx) +  \phiex(x) \,,\\
	\phiex(x) &= \sum_\ell \left[ a_\ell(t) u_\ell(\bx) + a_\ell^\d(t) v_\ell^*(\bx) \right] \,.
\end{align}
The first two terms in Eq.~(\ref{eq:phi_expansion}) correspond to the zero mode part, while $\phiex$
represents the excited modes. The commutation 
relation of $\phi$ leads us to
\be
	[Q(t), P(t)] = i \,,\quad
	[a_\ell(t), a_{\ell'}^\d(t)] = \delta_{\ell\ell'} \,,
\ee
and the vanishing ones otherwise, where $Q(t)$ and $P(t)$ are hermitian.
Substituting the expansion (\ref{eq:phi_expansion}) to Eq.~(\ref{eq:defH2}), we obtain
\be \label{eq:H2}
	H_2 = \frac{IP^2}{2} + \sum_\ell \omega_\ell a_\ell^\d a_\ell \,,
\ee
where $I = \partial\mu/\partial N_0$.
Thus the unperturbed Hamiltonian is diagonalized except for the zero mode part, which
involves the fatal dilemma in choosing 
the vacuum. If one chooses the zero momentum state with the least eigenvalue as the vacuum, 
one has $\avg{P^2(t)} = 0\,$, and the uncertainty relation implies $\avg{Q^2(t)} = \infty\,$, 
which is inconsistent with the assumption of the small $\phi$. In general, 
one may choose a wave packet state with finite $\avg{P^2(0)}\,$ at $t=0$ as the vacuum, 
but the Heisenberg equation
gives $Q(t)=Q(0)+ I P(0) t$ and $\avg{Q^2(t)}$ grows as $t^2\,$.
It shows that the choice of the wave packet is inadequate, or that it is valid only for a short time. 
Moreover, note that the unperturbed total atom number $\intx \!\avg{\psi^\d\psi}$ also grows as $t^2\,$.
To avoid the difficulties, Lewenstein and You have introduced a new expression of $\phi(x)$ \cite{Lewenstein},
\be  \label{eq:LewensteinAnsatz}
	\psi_{\rm LY}  \equiv \bigl( \xi + \eta P + \phiex \bigr)e^{-iQ} 
	\simeq \xi  -iQ\xi  + P \eta + \phiex \,,
\ee
where the last approximate expression is true only for small $Q$. Then  the conservation
of the total atomic number is recovered for the short time duration. However, 
we emphasize that because $\psi_{\rm LY}$ violates the canonical commutation relations which is 
the foundation of the quantum field theory, the formulation of quantum field theory in the interaction
picture as a whole becomes unfounded.

\section{Treatment of Zero Mode Beyond Bogoliubov Approximation}
The discussions above indicate that the simultaneous assumptions of 
the linear expansion of $\phi$, the bilinear unperturbed Hamiltonian  
and small $\phi$, are incompatible in treating the 
inevitable zero mode. We lift the choice of the unperturbed Hamiltonian, 
keeping the linear expansion and small $\phi$, and instead 
include the terms with the third and forth order powers
 of the zero mode operators into the unperturbed Hamiltonian $H_u$ as follows: 
\begin{align} \label{eq:Hu}
	H_u = H_1 + H_2 + \left[H_3 + H_4\right]_{QP} -\delta\mu P \,,
\end{align}
where the symbol $\left[ \cdots \right]_{QP}$ represents to pick up all the terms consisting only of 
the zero mode operators $Q$ and $P$, and the 
coefficient of the counter term $\delta\mu$ will be determined later.
The remaining terms, {\it e.g.\/} the terms such as $a^4$ or $P a^\d a$, are put into the interaction
Hamiltonian. It is stressed that the canonical commutation relations are respected because the temporal
evolution of $Q(t)$ and $P(t)$ are unitary, such as $Q(t)= e^{i H_u t} Q(0) e^{-i H_u t}\,$.

Substituting the expansion (\ref{eq:phi_expansion}) into Eq.~(\ref{eq:Hu}), we gather it
as
$
	H_u = H_u^{QP} + H_u^{\mathrm{ex}} \,
$
with
\begin{align}
	H_u^{QP} &=  (2J-\mu+2B-\delta\mu - 4C)P + \frac{I-4D}{2}P^2  \notag\\
	&\hspace{0.5cm}+ 2BQPQ + 2DP^3 + \frac{1}{2}AQ^4 -2BQ^2 \notag\\&\hspace{1cm}+ CQP^2Q +\frac{1}{2}EP^4 \,,\\
	H_u^{\mathrm{ex}} &=  \intx (\phiex+\phiex^\d)(h_0-\mu+g\xi^2)\xi + \sum_\ell \omega_\ell a_\ell^\d a_\ell \,,
\end{align}
where
\begin{alignat}{3} \label{eq:defAI}
	A &= g\intx \xi^4 \,,\;\;&
	B &= g\intx \xi^3 \eta\,,\;\;&
	C &= g\intx \xi^2 \eta^2\,,\;\; \notag\\
	D &= g\intx \xi \eta^3 \,,\;\;&
	E &= g\intx \eta^4 \,,\;\;&\notag\\
	I &= \pfrac{\mu}{N_0} \,,\;\;&
	J &= \intx \eta h_0 \xi \,.&
\end{alignat}
Since $H_u$ contains no cross-term between $\{Q,P\}$ and $\phiex$, 
the whole unperturbed vacuum $\ket0$ is expressed as the direct product
$\ket0= \ket\Psi \otimes \ket{0}_{\mathrm{ex}}$
where $\ket\Psi$ and $\ket{0}_{\mathrm{ex}}$ are vacua of the zero mode and excited mode sectors,
respectively. 
The criterion of division $\bra0 \phi \ket0=0$ leads 
\be \label{eq:criterion_of_division}
	\brasub_{\mathrm{ex}}\bra0 \phiex \ket{0}_\mathrm{ex} = 0 \,,\quad
	\bra\Psi Q \ket\Psi = 0 \,,\quad	
	\bra\Psi P \ket\Psi = 0 \,.
\ee
The time derivative of the first equality, in the same manner as in Eq.~(\ref{eq:timederivation}), 
derives the GP equation (\ref{eq:GP}), which implies $2J-\mu+2B=0$\,. 
The zero mode part of the unperturbed Hamiltonian $H_u^{QP}$ then becomes
\begin{align} \label{eq:HuQP}
	H_u^{QP} &=  -(\delta\mu + 4C)P + \frac{I-4D}{2}P^2 + 2BQPQ + 2DP^3  \notag\\
	&+ \frac{1}{2}AQ^4 -2BQ^2 + CQP^2Q +\frac{1}{2}EP^4 \,.
\end{align}
Similarly, the time derivative of the second equality in Eq.~(\ref{eq:criterion_of_division}) 
derives the identity
\begin{align}
	&\bra\Psi \bigl[-\delta\mu - 4C + (I-4D)P + 2BQ^2  \notag\\&\hspace{1.5cm} +6DP^2
	+ 2CQPQ + 2EP^3 \bigr]\ket\Psi = 0 \,,
\end{align}
which fixes $\delta\mu$ as 
\begin{align}\label{eq:deltamu}
	&\delta \mu = \bra\Psi\bigl[ - 4C + 2BQ^2 +6DP^2 + 2CQPQ + 2EP^3  \bigr]\ket\Psi \,,
\end{align}
to satisfy the third equality in Eq.~(\ref{eq:criterion_of_division}).
Here, the vacuum of the zero mode $\ket\Psi$ should be the ground state of the stationary Schr\"odinger equation,
\be \label{eq:HuEigen}
H_u^{QP} \ket\Psi= E_0\ket\Psi \,,
\ee
and $\delta\mu$ is to be determined self-consistently.
As there are terms with odd powers of $P$ but not of $Q$ in $H_u^{QP}$, 
the second equality in Eq.~(\ref{eq:criterion_of_division}) is satisfied automatically
when $\braket{q}{\Psi}$ is an even function of $q$.

\subsection{Variational Estimation for Homogeneous System}
We first estimate $\ket\Psi$ variationally with the trial function,
\be \label{eq:PsiVar}
	\braket{q}{\Psi} = \left(\frac{1}{2\pi\alpha^2}\right)^{1/4} e^{-\frac{q^2}{4\alpha^2}}\,.
\ee
The variational parameter $\alpha$ is related to the expectation value of $Q^2$ as $\alpha^2 =  \bra\Psi Q^2 \ket\Psi$.
From Eq.~(\ref{eq:deltamu}), $\delta\mu$ is also expressed in terms of $\alpha$ as $\delta\mu = 2B\alpha^2 - 4C + 3D/2\alpha^2\,$. 
The parameter $\alpha$ is determined to minimize the expectation value $f(\alpha) = \bra\Psi H_u^{QP} \ket\Psi\,$, that is 
\be \label{eq:dfda=0}
	\pfrac{f}{\alpha} = 6A\alpha^3 -4B\alpha^2 - \frac{I-4D}{4\alpha^3}- \frac{3E}{8\alpha^5} = 0 \,.
\ee
In the large $N_0$ limit, only the terms proportional to $A$ and $I$ are dominant in Eq.~(\ref{eq:dfda=0}), and we obtain 
\be \label{eq:alpha}
	\alpha = \sqrt[6]{\frac{I}{24A}} \,.
\ee
When we consider a homogeneous system ($V_{\mathrm{ex}}=0$ and $\xi=$ constant) 
where $\xi = N_0/V \,, \mu = gN_0/V \,,$ and $I=g/V$ with the volume $V$, 
Eq.~(\ref{eq:alpha}) leads to $\alpha = (1/{\sqrt[6]{24}})\, N_0^{-1/3}$. It is independent of both 
$g$ and $V$, and depends only on $N_0$. The standard deviation of $Q$, denoted by 
$\Delta Q= \sqrt{\bra\Psi Q^2 \ket\Psi-\bra\Psi Q {\ket\Psi}^2}$, is equal 
to $\alpha$ since $\bra{\Psi} Q \ket{\Psi}=0$ for the trial function (\ref{eq:PsiVar}).
As long as $\Delta Q$ is sufficiently small, it is interpreted as the fluctuation of the phase,
$\psi \simeq \xi -i  Q \xi + \cdots \simeq e^{-i  Q} \xi+\cdots$, similarly as in
 Eq.~(\ref{eq:LewensteinAnsatz}) which is not adopted in our approach though.
Our estimation of Eq.~(\ref{eq:alpha}) shows that the 
uncertainty of the condensate phase decreases as $N_0^{-1/3}$, and vanishes at the thermodynamical limit. 

\begin{figure}[tb!]
\begin{center}
\vspace{1cm}
\includegraphics[width=7cm]{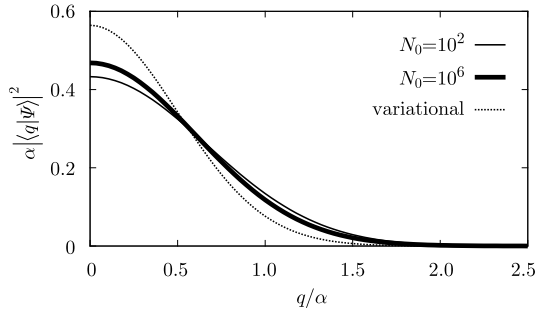}
\end{center}
\caption{\footnotesize{%
The ground state distribution for the homogeneous system. The thin and bold lines denote the numerical 
results for $N_0=10^2$ and $10^6$, respectively. For comparison the variational result is shown
as the dotted line. 
As the axes are scaled by $\alpha$, the dotted line is true for an arbitrary $N_0$.
}}
\label{figPsi}
\end{figure}
\begin{figure}[tb!]
\begin{center}
\vspace{1cm}
\includegraphics[width=7cm]{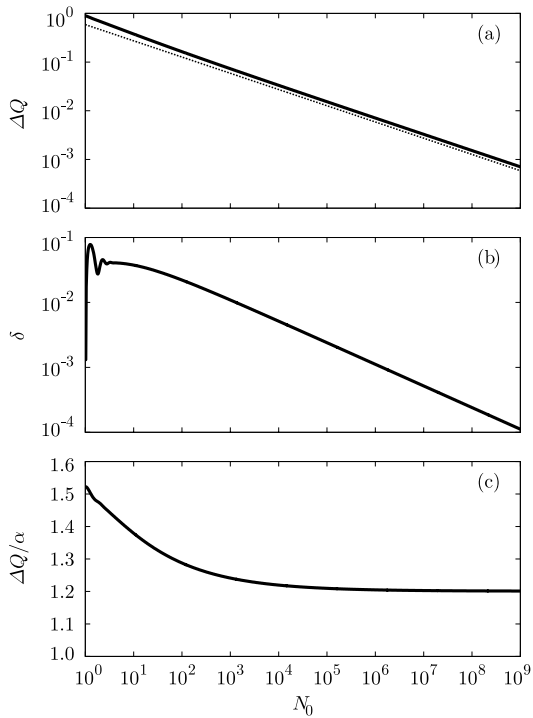}
\end{center}
\caption{\footnotesize{%
$N_0$ dependences of (a) $\Delta Q$, (b) $\delta$, and (c) $\Delta Q/\alpha$ for the homogeneous system.
The dotted line in (a) indicates the variational estimation $\alpha = (1/\sqrt[6]{24}) N_0^{-1/3}$. 
The quality $\delta = -1/3- \frac{d\log\Delta Q}{d \log N_0}$ in (b) indicates the shift of the 
exponent of $\Delta Q$ from $-1/3$.
}}
\label{fig}
\end{figure}

\subsection{Numerical Calculation for Homogeneous System}
We next solve  Eq.~(\ref{eq:HuEigen}) numerically for the homogeneous system.
The numerical results of the ground state distribution is depicted in Fig.~\ref{figPsi},
and allow us to calculate $\Delta Q\,$. 
Figure~\ref{fig}~(a) shows that the variational estimation ($\alpha$, dotted line) is in good 
agreement with the numerical result ($\Delta Q$, solid line). Let us introduce the quantity $\delta$ by
parameterizing $\Delta Q  \sim N_0^{-(1/3+\delta)}$.  As shown in Fig.~\ref{fig}~(b), 
$\delta$ decreases exponentially for large $N_0$
and implies that the exponent converges to $-1/3$ in the limit $N_0\to\infty\,$, 
as is estimated variationally. Unlike in the case of the exponent, 
the coefficient in $\Delta Q$ differs from the variational estimation, and the ratio $\Delta Q/\alpha$ 
converges to about $1.20$ as Fig.~\ref{fig}~(c) shows. It is not negligible small but one may
conclude that the variational estimation is valid.

\subsection{Inhomogeneous System}
Finally, we confirm that the behavior of $\Delta Q \sim N_0^{-1/3}$ for large $N_0$ is universal even for an 
inhomogeneous system with $V_{\mathrm{ex}}(\bx)\neq 0$ and  is independent of the interaction strength, 
dimension, and the shape of the confinement potential.
We consider a $d$-dimensional system with a repulsive interaction $g>0\,$, 
and the attractive case is excluded because then the condensate collapses for large $N_0$. 
Let us suppose that $V_{\mathrm{ex}}(\bx)$ has a form of homogeneous
function: $V_{\mathrm{ex}}( \lambda\bx) = \lambda^\nu V_{\mathrm{ex}}(\bx)$ with a parameter $\nu\geq 1$.
 Then the virial theorem yields \cite{PethickSmith} the relation
$2E_{\mathrm{kin}} -\nu E_{\mathrm{pot}} + dE_{\mathrm{int}} =0$, where 
$E_{\mathrm{kin}} = \frac{1}{2m}\int \!d^dx \, \left| \nabla \xi(\bx)\right|^2$\,,
$E_{\mathrm{pot}} = \int \!d^dx \, V_{\mathrm{ex}}(\bx) \left|\xi(\bx)\right|^2$\,, and
$E_{\mathrm{int}} = \frac{g}{2} \int \!d^dx \, \left|\xi(\bx)\right|^4$\,. 
Besides, one derives $E_{\mathrm{kin}} +E_{\mathrm{pot}} + 2 E_{\mathrm{int}}
 =\mu N_0$ from the Gross-Pitaevskii equation.
Taking the large $N_0$ limit and using the Thomas--Fermi approximation, $E_{\mathrm{kin}}=0$, we obtain
$\mu= E_{\mathrm{int}} (d+2\nu)/\nu N_0$. 
The interaction energy $E_{\mathrm{int}}$ is equal to $A/2$ [see Eq.(\ref{eq:defAI})],
 and its leading term in 
the large $N_0$ limit can be expressed as $A\simeq c_1N_0^{c_2}$ with $c_1>0$ and $c_2>1$. 
Using Eqs.~(\ref{eq:defAI}) and 
(\ref{eq:alpha}), we obtain
\be
	\Delta Q =\sqrt[6]{\frac{(d+2\nu)(c_2-1)}{48\nu}} N_0^{-1/3}\,,
\ee
where the power of $N_0$ is independent of the other parameters and equals to $-1/3$. 

\SEC{Summary}
In summary, including the terms with higher powers of the zero mode operators $Q$ and $P$ 
in the unperturbed Hamiltonian, 
we have obtained the appropriate vacuum 
which is stationary and whose validity is not restricted to a finite time duration.
This is a contrast to the previous theoretical formulation in which
no discrete ground state exist and the phase of the order parameter diffuses out.
The difference comes simply from the choices of the unperturbed Hamiltonian in 
both the cases. We have evaluated the standard variation $\Delta Q$ 
which is interpreted as the condensate phase when $\Delta Q\ll 1$. The phase fluctuation $\Delta Q$ decreases as 
$\sim N_0^{-1/3}$ for large $N_0$. This power law is universal and independent of
the other parameters for the both homogeneous and inhomogeneous systems. 
We note that while the power law is independent of the interaction constant $g$, 
the presence of non-zero $g$ is crucial to the robustness of the condensate phase against the quantum fluctuation of the 
zero mode. 
If the higher powers of the zero mode operators, {\it i.e.} $[H_3+H_4]_{QP}$,  
are excluded from the unperturbed Hamiltonian, 
the present formulation reduces to that of Lewenstein and You \cite{Lewenstein}. 
The higher power contribution of the zero mode operators is essential to go beyond Bogoliubov approximation 
and to determine a unique stationary vacuum.
In this paper, we have considered only the case 
where the quantum fluctuation of the exited modes and the thermal one of all the modes are negligibly small.
To extend the present formulation to a system at finite temperature would be the future work.

\begin{acknowledgments}
This work is partly supported by Grant-in-Aid for Scientific Research (C) (No. 25400410) from the Japan Society for the 
Promotion of Science, Japan; ``Ambient SoC Global Program of Waseda University" of the Ministry of Education, Culture, 
Sports, Science and Technology, Japan; and Waseda University Grant for Special Research Projects (Project No. 2013B-102). 
\end{acknowledgments}

\end{document}